\documentclass[10pt,conference]{IEEEtran}
\IEEEoverridecommandlockouts
\usepackage{cite}
\usepackage{amsmath,amssymb,amsfonts}
\usepackage{algorithmic}
\usepackage{graphicx}
\usepackage{textcomp}
\usepackage{xcolor}
\usepackage{blindtext}
\usepackage{algorithm}
\usepackage{comment}
\usepackage{url}

\usepackage{bytefield}
\usepackage{caption}
\usepackage[colorinlistoftodos,prependcaption]{todonotes}

\def\BibTeX{{\rm B\kern-.05em{\sc i\kern-.025em b}\kern-.08em
    T\kern-.1667em\lower.7ex\hbox{E}\kern-.125emX}}
\begin{document}

\title{QUIC-FEC: Bringing the benefits of Forward Erasure Correction to QUIC}

\author{\IEEEauthorblockN{Fran{\c c}ois Michel}
\IEEEauthorblockA{\textit{UCLouvain} \\
Louvain-la-Neuve, Belgium\\
francois.michel@uclouvain.be}
\IEEEmembership{FNRS Research Fellow}
\and
\IEEEauthorblockN{Quentin De Coninck}
\IEEEauthorblockA{\textit{UCLouvain} \\
Louvain-la-Neuve, Belgium\\
quentin.deconinck@uclouvain.be}
\IEEEmembership{FNRS Research Fellow}
\and
\IEEEauthorblockN{Olivier Bonaventure}
\IEEEauthorblockA{\textit{UCLouvain} \\
Louvain-la-Neuve, Belgium\\
olivier.bonaventure@uclouvain.be}
}

\maketitle

\begin{abstract}
Originally implemented by Google, QUIC gathers a growing interest by
providing, on top of UDP, the same service as the classical
TCP/TLS/HTTP/2 stack. The IETF will finalise the QUIC specification in 2019.

A key feature of QUIC is that almost all its packets, including most
of its headers, are fully encrypted. This prevents eavesdropping and
interferences caused by middleboxes. Thanks to this feature and its
clean design, QUIC is easier to extend than TCP.
In this paper, we revisit the reliable transmission mechanisms
that are included in QUIC. More specifically, we design, implement and
evaluate Forward Erasure Correction (FEC)
extensions to QUIC. These extensions are mainly intended for high-delays and lossy
communications such as In-Flight Communications. Our design includes a generic FEC frame and our implementation supports the XOR, Reed-Solomon and
Convolutional RLC error-correcting codes. We also conservatively avoid
hindering the loss-based congestion signal by distinguishing the
packets that have been received from the packets that have been
recovered by the FEC.
We evaluate its performance by applying an experimental
design covering a wide range of delay and packet loss conditions with
reproducible experiments. These confirm that our modular
design allows the protocol to adapt to the network conditions. For long data transfers or
when the loss rate and delay are small, the FEC overhead negatively
impacts the download completion time. However, with high packet loss
rates and long delays or smaller files, FEC allows drastically
reducing the download completion time by avoiding costly retransmission timeouts. These results show that there is a need to use FEC adaptively to the network conditions.

\end{abstract}


\section{Introduction}

Initially proposed by Google engineers to reduce web page download times, the
QUIC protocol \cite{langley2017quic} brought innovation back in the transport layer. QUIC started as an evolution of SPDY \cite{spdy}, a precursor of HTTP/2. In a nutshell, QUIC combines in a single protocol atop UDP the mechanisms that are usually found in three different protocols:  TCP, TLS and HTTP/2. In contrast with TLS/TCP, QUIC encrypts both the payload and most of the protocol control information to prevent both pervasive monitoring and ossification from
middleboxes. Since it is built over UDP, QUIC is easier to update than TCP. Indeed, QUIC implementations can be easily included as libraries inside applications that are regularly updated. Recent measurements show that a growing number of servers (mainly Google and Akamai) support Google's version of QUIC~\cite{ruth2018first} but also that the QUIC traffic grows~\cite{langley2017quic,Trevisan_Edge:2018}.

Given the positive results obtained by Google with QUIC \cite{langley2017quic}, the IETF created a dedicated working group in 2016 to standardise a new protocol starting from
Google's initial design \cite{quic-draft-00}. %
The first stable QUIC specification is expected by the end of July 2019 and more than a dozen implementations are being developed. Although the initial use-case will be HTTP/2 \cite{quic-http}, QUIC could also be used to support DNS~\cite{dnsoquic}, RTP
\cite{Ott_RTP-QUIC:2017} and unreliable messages \cite{Pauly_unreliable:2018}.

One of the strengths of the QUIC protocol is its extensibility. A QUIC packet payload contains a sequence of \textit{frames}, each one being handled independently by the protocol. The \textit{Stream frame} transports application data. The \textit{ACK frame} acknowledges the received packets to the sender. New types of frames can easily be added to the protocol. Furthermore, since QUIC packets are encrypted, middleboxes cannot interfere with protocol extensions, in contrast with the problems that have plagued TCP \cite{honda2011still}.

In this article, we extend QUIC by enabling it to rely on Forward Erasure Correction (FEC) to recover from packet losses. This design is motivated by high Bandwidth-Delay Product (BDP) networks such as In-Flight Communication (IFC) services where losses are frequent and retransmissions impact user experience \cite{inflightconnectivity}.  These techniques transmit redundant code to enable the receiver to recover from packet losses without waiting for retransmissions. FEC techniques have already been used in multicast applications \cite{carle1997survey,perkins1998survey} or in
TCP \cite{mptcpfec,fmtcp,sundararajan2011network, cloud2013multi}, but the TCP extensions are difficult to deploy~\cite{honda2011still}. %

We propose three main contributions in this paper. First, we propose a modular QUIC extension that enables the utilisation of a variety of FEC techniques. Our extension goes much beyond the experiments carried out by Google with a simplistic FEC technique in Chrome \cite{quicfecprague}. %
Furthermore, our design makes the congestion control aware of packets that were either normally received or recovered by FEC.
Second, we provide a complete implementation of the proposed extension in \texttt{quic-go} \cite{quic-go} with three different FEC techniques. Third, our evaluation, over a wide range of parameters, indicates that the proposed FEC techniques improve the performance of QUIC for short file transfers.

This paper is organised as follows. We first describe the concepts behind FEC and discuss its support within QUIC in Section \ref{section_fec}. We then define in Section \ref{section_design_implem} the design and implementation details of QUIC-FEC, our extension enabling the use of FEC-protected transfers with QUIC. We finally assess its performances and compare different erasure-correcting codes through experiments using a wide range of network and loss configurations in Section \ref{section_evaluation}.

\section{Forward Erasure Correction}\label{section_fec}

Over the last decades, researchers have explored a variety of techniques that transmit redundant data to enable the receiver to recover from errors and losses without having to wait for retransmissions. 
Some proposed techniques were tuned for specific link layer technologies \cite{biersack1993performance,katti2008xors,giesen2018network} or targeted for specific applications \cite{carle1997survey,perkins1998survey,sundararajan2011network,fmtcp,mptcpfec,cloud2013multi}. 
The IETF also considers these techniques within the RMT and FECFRAME working groups and the IRTF NWCRG \cite{Adamson_Taxonomy:2018}. FEC is especially interesting compared to retransmission mechanisms when the delay and loss rate are high: the packets will be recovered by FEC without the need for a retransmission. Rula \textit{et al.}~\cite{inflightconnectivity} recently revealed that In-Flight Communications (IFC) are highly deteriorated by the important latency and loss rate, making it an interesting candidate for evaluating the benefits of FEC.
IFC technologies rely on 3/4G and satellite technologies. 
Despite built-in redundancy and retransmission mechanisms often proposed by such technologies, they may not be able to recover from transmission losses, especially when the user is mobile \cite{draft-network-coding-satellites}, which explains why losses can be perceived from higher layer perspectives in the IFC use-case. 
There is thus a call for exploiting coding schemes in higher OSI layers.

In their work on IFC communications, Rula \textit{et al.}~\cite{inflightconnectivity} study the potential impact of the new arriving technologies for IFC aiming at improving the link's bandwidth. 
They conclude that improving the link bandwidth does not improve significantly the performances as the bottleneck resides in the high losses and latencies. 
They also recognise that reducing the latency and loss rate in this use-case is challenging.
We thus study the impact of adding FEC in QUIC for IFC, as FEC might benefit from having more bandwidth provided by these new technologies to reduce the impact of losses during a connection over a lossy channel. 

Given that most link layer technologies include error detection codes, we focus on erasures instead of errors. We define FEC as the transmission of redundancy -- Repair Symbols -- along with the data to recover packets -- Source Symbols -- that have been \textit{lost}. 
We use the word \textit{FEC scheme} to refer to the handling of Source and Repair Symbols and their redundancy generation using an erasure correcting code.

\subsection{Current support of FEC in QUIC}

While a simple FEC scheme was originally included in the QUIC protocol~\cite{quic-draft-00}, it has rapidly been dropped due to negative experiments results~\cite{quicfecprague}.
However, these experiments targeted classical web use-cases where the loss rate and delays are quite low compared to other use-cases such as In-Flight Communications that have not been explored.

Furthermore, the FEC scheme used by Google was simplistic, being only able to recover single losses, while bursty losses occur in practice~\cite{quicfecprague}.
A wide range of codes more adapted to those network conditions have been proposed in the literature~\cite{perkins1998survey}.
In this work, we consider both block and convolutional error-correcting codes.

\begin{figure}
\centering
\includegraphics[width=.9\linewidth]{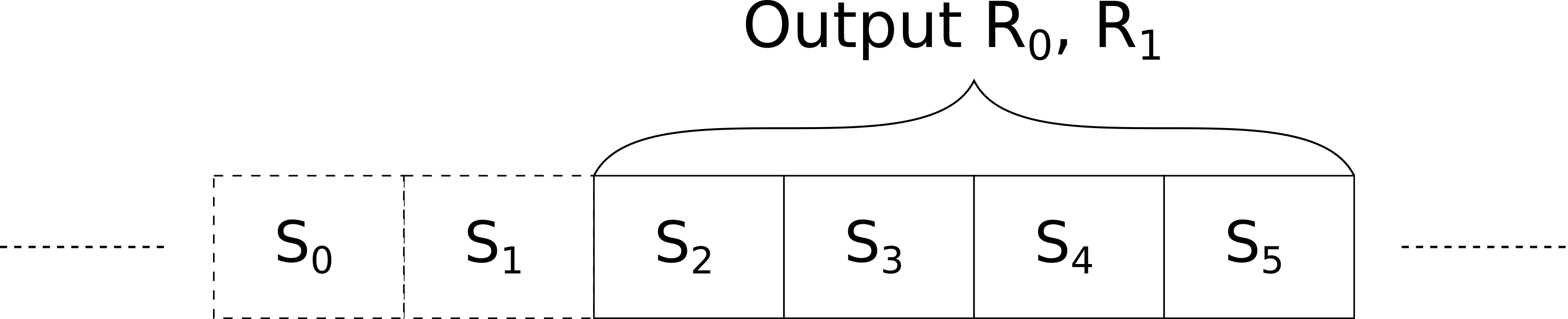}
\caption{Example of $(6, 4)$ block code. 2 Repair Symbols are send for each block of 4 distinct Source Symbols.}
\label{figure_block_code_example}
\end{figure}

\begin{figure}
\centering
\includegraphics[width=.9\linewidth]{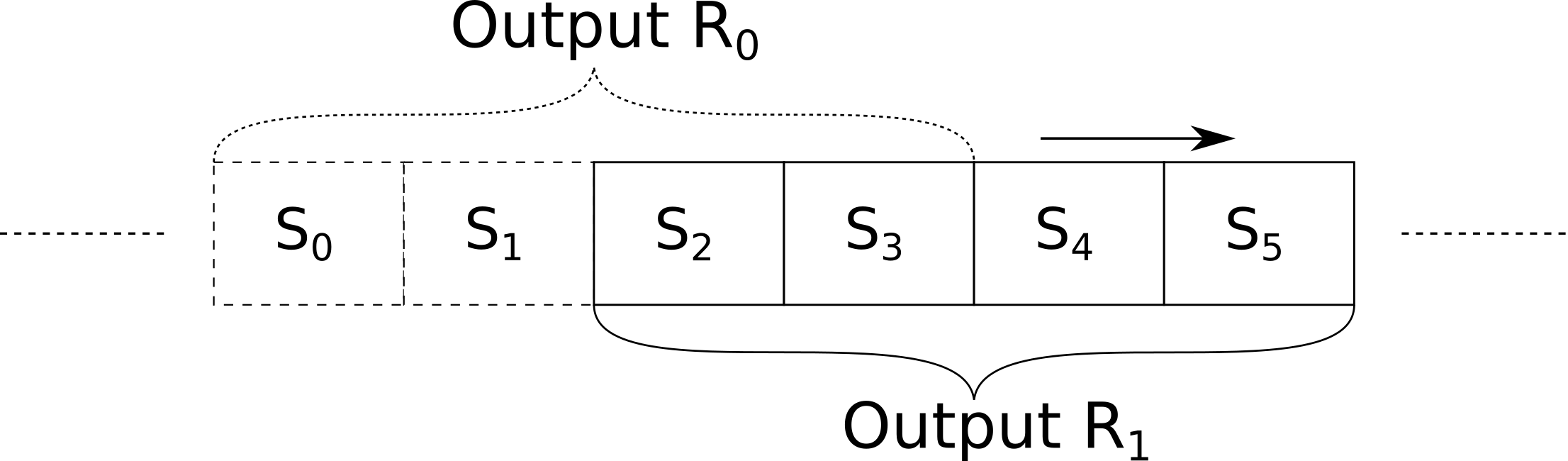}
\caption{Example of $(3, 2, 4)$ convolutional code. A coding window of 4 Source Symbols slides with a step of 2 symbols. For each window position, it outputs 1 Repair Symbol protecting the 4 Source Symbols contained in the current window.}
\label{figure_convo_code_example}
\end{figure}

We define a $(n, k)$ block code as a code sending a block of $k$ Source Symbols followed by $n-k$ Repair Symbols. Figure \ref{figure_block_code_example} shows an example of a $(6, 4)$ code: each block of 4 Source Symbols is protected by 2 Repair Symbols (here, $R_0$ and $R_1$). We define a $(n, k, c)$ convolutional code as a convolutional code sending $n-k$ Repair Symbols every $k$ Source Symbols. 
The Repair Symbols protect the $c$ previous Source Symbols. Figure \ref{figure_convo_code_example} shows an example of a $(3, 2, 4)$ convolutional code. It sends one Repair Symbol every two Source Symbols. 
This means that the Repair Symbols are sent interleaved with the Source Symbols. 
Each Repair Symbol protects the 4 previously sent Source Symbols.

From the viewpoint of the QUIC protocol, the packet reception outcome is binary. Either the packet is fully received or it is not received at all. 
We thus consider entire packets as Source Symbols, instead of protecting an arbitrary number of bits in the packets. This is called \textit{packet-level coding}.

Roca \textit{et al.\ } compare block and convolutional codes using Reed-Solomon and Random Linear Codes (RLC) to represent both families of codes \cite{roca2017less}. 
They show that while the Reed-Solomon block codes provide a higher encoding speed, RLC allows recovering the packets with a reduced latency compared to Reed-Solomon. We leverage this property within our QUIC extension. Palmer \textit{et al.}~\cite{palmer2018quic} recently studied the combination of an unreliable transfer and FEC on top of QUIC for Video-on-Demand and live video streaming. However, only one block code has been considered and using an unreliable transfer for these use-cases could have been avoided by using a sufficient playback buffer. To the best of our knowledge, the video conferencing use-case using QUIC and FEC has not been discussed yet.

\section{Integrating FEC into QUIC} \label{section_design_implem}

In this section, we propose a generic Forward Erasure Correction extension for QUIC. It is currently based on the \textit{Google} version of QUIC and we implement it using the \texttt{quic-go} implementation. 
We have a similar design for the \textit{IETF} version of QUIC \cite{quic-draft-16} implemented in \texttt{picoquic} \cite{picoquic}, but do not discuss it due to space limitations. Our design transparently supports different FEC Schemes and allows easily adapting the selected FEC Scheme. 
The application can select the FEC Scheme that suits its needs and recover from losses without waiting for retransmissions. 

Adding such a mechanism requires addressing several points. We first describe how we advertise which are the Source and Repair Symbols to the peer in Section \ref{section_symbols}. We then explain how we manage to transparently handle different FEC Schemes in Section \ref{section_fec_framework}. We finally discuss the impact of FEC on congestion control in Section \ref{section_fec_congestion}.

\subsection{Sending the Source and Repair Symbols} \label{section_symbols}
Our design uses QUIC packets as Source Symbols. 
We use a previously unused flag of the QUIC header to inform the peer that the packet must be considered as a Source Symbol. 
We add a 32-bits field to the packet header to transmit FEC Scheme-specific values to the peer.
We only protect the packets containing Stream frames that carry user data. Successive ACK frames contain redundant information, making their loss less impacting than the loss of Stream frames.

\begin{figure}[t]
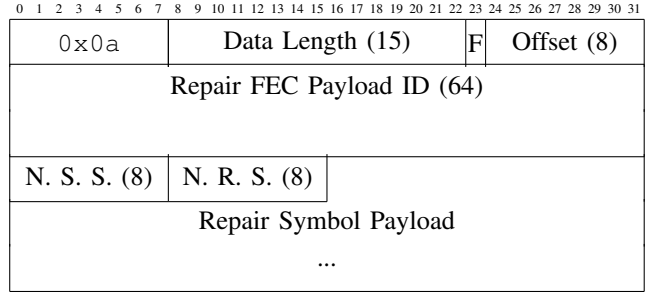

\centering
\begin{bytefield}[bitwidth=.75em]{32}
\bitheader{0-31} \\
\bitbox[ltr]{8}{\texttt{0x0a}}
\bitbox[ltr]{15}{Data Length (15)}
\bitbox[tr]{1}{F}
\bitbox[tr]{8}{Offset (8)} \\
\bitbox[ltr]{32}{Repair FEC Payload ID (64)} \\ 
\bitbox[lrb]{32}{} \\
\bitbox[lrb]{8}{N. S. S. (8)}
\bitbox[rb]{8}{N. R. S. (8)}
\bitbox[r]{16}{} \\
\bitbox[lr]{32}{Repair Symbol Payload} \\
\bitbox[lrb]{32}{...} \\
\end{bytefield}
\caption{Wire format of a FEC frame. The Repair FEC Payload ID field is opaque to the protocol and is populated by the underlying FEC Scheme.}
\label{gquic_fec_frame_format}
\end{figure}

We introduce a new QUIC frame, the \textit{FEC frame}, depicted in Figure \ref{gquic_fec_frame_format}. The FEC frame contains the Repair Symbols payload as well as a 64-bits (\textit{Repair FEC Payload ID)} field containing FEC Scheme-specific values. The FEC Scheme-specific values are handled by the underlying FEC Scheme and are opaque to the QUIC protocol itself. It allows easily developing new FEC Schemes independently of the behaviour and core functionalities of the QUIC protocol. The \textit{FEC frame} also advertises the number of Source (\textit{N.S.S.}) and Repair Symbols (\textit{N.R.S.}) in the current FEC Block for block codes. For convolutional codes, these fields advertise the number of Source Symbols in the sliding window and the number of Repair Symbols generated at each window step. They allow the sender to dynamically change the code rate during the connection and adapt its use of FEC to the changing network conditions.

As a single Repair Symbol could be too large to fit into a single FEC frame, the latter contains an \textit{Offset} field indicating the offset of the Repair Symbol chunk transported in the frame and a FIN bit (\textit{F}) indicating if the frame contains the last chunk for this Repair Symbol.

While the QUIC Packets are sent encrypted and authenticated, the Repair Symbols are generated from their clear-text, avoiding the CPU overhead of decrypting the recovered packets. To ensure the confidentiality and integrity of the recovered packets, the Repair Symbols are sent encrypted and authenticated. Therefore, no FEC frame is sent before the end of the cryptographic handshake.

\subsection{The FEC Framework}\label{section_fec_framework}
The IETF has already developed solutions to add error-correcting codes to several protocols. The most recent solution
is the FECFRAME framework \cite{fecframe} which has notably been
applied to RTP and supports different FEC schemes 
\cite{reedsolomonfecframe,rfc6816}. 

Inspired by FECFRAME \cite{fecframe}, we define a FEC Framework implementing the common behaviour of these FEC Schemes in order to further simplify their implementation. It provides a structure for the different FEC Scheme-specific values exchanged by the peers to proceed to erasure correction.  Its design is described in details in a technical report~\cite{quicfec12pages}.
Although there exists a wide variety of FEC Schemes, we can classify most of them in two main categories: FEC Schemes using block codes and FEC Schemes using convolutional (a.k.a. sliding window) codes. Our FEC Framework is designed to handle both. For block codes, it uses the 32-bits field in the packet header to encode the FEC block number and the offset of this packet in the FEC block. For convolutional FEC Schemes, it uses these 32 bits to encode  the offset of this packet in the protected packets sequence. The framework also populates 32 bits of the \textit{Repair FEC Payload ID} field in the FEC frame with informations identifying the FEC Block (for block codes) and coding window (for convolutional codes) protected by the Repair Symbol. The 32 other bits can be populated by the underlying FEC Scheme with values required to perform the encoding/decoding.

Such interfacing also brings interest from the IETF, where the network coding research group currently works on an Internet Draft~\cite{coding-for-quic}.
While the core ideas are similar, the draft recommends to only protect stream chunks while our implementation protects arbitrary QUIC frames of any type.
We believe our work will benefit to the standardisation of such interfacing.

\textbf{Studied FEC Schemes.} We currently support three different FEC Schemes, each of them having different characteristics: the XOR, the Reed-Solomon and the Convolutional Random Linear Code (RLC) FEC Schemes. The two first ones are block FEC Schemes and the last one is a convolutional FEC Scheme.

\textit{XOR FEC Scheme.} Its principle is quite simple: the Source Symbols are simply XORed with each other to generate a Repair Symbol. It is easy to implement and to compute but can only recover the loss of one Source Symbol. Experiments carried out
by Google showed that this is insufficient on the Internet \cite{Swett_FEC:2016} because losses can occur in bursts. Our implementation uses interleaving to recover from burst losses with the XOR FEC Scheme.
Sending successive packets in different FEC Blocks enables the XOR FEC Schemes to better handle burst losses at the expense of delay.

\textit{Reed-Solomon FEC Scheme.} It can generate multiple Repair Symbols per Source Block, allowing handling of burst losses. While it better handles burst losses than the XOR FEC Scheme, it is also more computationally intensive.

\textit{Convolutional RLC FEC Scheme.} As convolutional codes provide different properties from block codes, our FEC extension enables their use through the RLC FEC Scheme, solving a system of linear equations with the lost Source Symbols. It allows interleaving the Repair Symbols along with the Source Symbols.
Our implementation is inspired from the \textit{FECFRAME} RLC FEC Scheme draft \cite{rlc-fecframe}. 

\subsection{FEC and the congestion control} \label{section_fec_congestion}

Different congestion control algorithms have been implemented into transport protocols such as QUIC and TCP. Cubic~\cite{ha2008cubic} and New Reno~\cite{henderson2012newreno} are the most popular ones. 
In the case of QUIC-FEC, there are three outcomes to a packet loss. $i)$ \textit{The packet was not FEC-protected.} In that case, it will not be recovered. The sender observes a hole in the acknowledgements and registers a loss. Packets containing only FEC frames fall in this category. $ii)$ \textit{The packet was FEC-protected but could not be recovered.} In that case, the sender will notice the loss and retransmit the missing Stream frames. $iii)$ \textit{The packet was FEC-protected and recovered.} Acknowledging these recovered packets can hide the congestion signal and make the FEC-enabled protocols behave unfairly compared to TCP or regular QUIC, as they could potentially take more than their fair share of the link bandwidth. We considered three ways of avoiding to hide the congestion notification signal due to packet losses:

\subsubsection{Not acknowledging the recovered packets} 
This approach conservatively considers a recovered packet as lost. It leads to a similar congestion control behaviour to when the lost packets are not recovered with FEC. The drawback is that the sender will perform unnecessary packet retransmissions. 
\subsubsection{Distinguishing the congestion-implied packet losses from the channel noise-implied losses} 
Kim \textit{et al.} \cite{kim2014congestion} propose to modify the loss-based congestion controls for this purpose. 
They assume that a congestion-implied loss event is preceded by an increase of the Round-Trip-Time (RTT) due to the filling of the network buffers. 
They propose to diminish the decrease of the congestion window after a packet loss if the current RTT is close to the minimum observed RTT. Tickoo \textit{et al.}~\cite{tickoo2005lt} propose to only react to congestion when it is explicitly notified by the network nodes though the Explicit Congestion Notification (ECN)~\cite{ramakrishnan2001addition} mechanism.
TCP Westwood \cite{mascolo2001tcp} estimates the link bandwidth using the acknowledged data rate. At each loss event, it adjusts its congestion window to use the estimated bandwidth instead of multiplicatively decreasing it.

\subsubsection{Explicitly advertising a packet recovery to the sender} This approach extends the conservative approach. The packet is acknowledged in a QUIC ACK frame but the receiver also signals that this packet has been recovered. Upon reception of this information, the sender both adapts its congestion window according to the loss event and removes the recovered packet from its retransmission queue. We implement this solution in QUIC-FEC with a \textit{Recovered frame}. Its format is similar to the QUIC ACK frame: it advertises the ranges of newly recovered packets. 

Once a sender receives a Recovered frame, it removes the recovered packets from its retransmission queue. It then signals to its congestion control that a packet has been lost for each packet listed in the Recovered frame. Using the Recovered frame conservatively adapts the congestion window of the sender as if every loss was caused by congestion. This behaviour is comparable to the utilization of ECN. Once a packet containing a Recovered frame is acknowledged, the recovered packets ranges are removed from the subsequent Recovered frames.
We analyse the impact of this approach in Section~\ref{subsection:recovered}.

\section{Evaluation} \label{section_evaluation}

Experiments have been
performed to assess the performances of our implementation and analyse the benefits of Forward Erasure Correction with the HTTP use-case running over QUIC. 
We first describe in this section our methodology, then perform experiments with parameters inspired by In-Flight Communications. We analyse the Download Completion Time (DCT), i.e. the time required to complete an HTTP transfer. 

\subsection{Methodology}

We use network emulation with the Mininet tool \cite{handigol2012reproducible} to evaluate the performance of different FEC schemes in \texttt{quic-go}. The
main benefit of using emulation is that it runs with real code and
not a simplified protocol model. We perform experiments with different loss models.
While small burst lengths or uniform losses can already give an idea on the efficiency of a solution, we advocate for looking at longer burst lengths as well in order to evaluate our solution with different loss configurations. We thus use the Gilbert-Elliott model \cite{elliottmodel}, a standard loss model used
to represent bursts of lost packets. 

The \textit{Gilbert-Elliott} model is a two-states Markov model used to represent correlated losses. 
The two states are the \textit{Good} and \textit{Bad} states. In the \textit{Good} state, a packet is delivered with a probability $k$. 
In the \textit{Bad} state, a packet is delivered with a probability \textit{h}. \textit{p} denotes the probability of transition from the \textit{Good} to the \textit{Bad} state, while \textit{r} denotes the probability of transition from the \textit{Bad} to the \textit{Good} state.

We added the packet-based loss detection mechanism to \texttt{quic-go} as described in the current IETF specification~\cite{quic-rec-draft-16} as the version of \texttt{quic-go} we use only supports a time-based loss detection marking a packet as lost when observing a hole in the acknowledged packet ranges for more than $\frac{1}{8}*RTT$.

\subsubsection{Experimental design}
\begin{figure}
\centering
\includegraphics[width=0.45\textwidth]{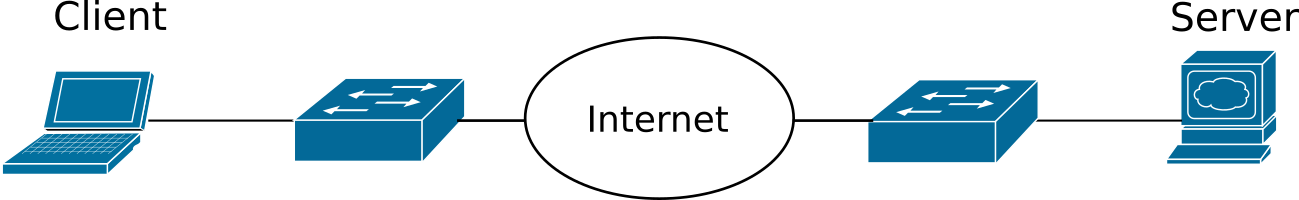}
\caption{Network topology for our experiments.}
\label{figure_exp_topo}
\end{figure}

We use an \textit{experimental design} approach to perform our experiments \cite{fisher1949design}. 
This methodology consists in defining ranges 
for each parameter and performing a series of experiments with well-chosen values within these ranges. 
This sub-samples the ranges of values and gives a global overview of the possible values taken by all the parameters. 
In addition to providing a general confidence concerning the performances of the tested implementation, it mitigates the bias in the parameters' selection by the experimenter. 
We use the WSP algorithm \cite{wsp} to broadly sample the space of parameters with a reasonable amount of experiments. Unless otherwise specified, we run the experiments with 130 different combinations of parameters.
Each configuration is run 9 times and the median download completion time from these 9 runs is considered. The experiment consists of an HTTP/2 GET request for a particular file using QUIC. Figure \ref{figure_exp_topo} shows the network topology used for our experiments. We apply the delay, losses and bandwidth limitation on the ``Internet'' link. 
Table \ref{table_parameters_range} shows the parameters ranges chosen for our experiments. It specifies ranges for the One-Way Delay (OWD), bandwidth (BW), uniform loss rate ($p$) when a uniform loss model is used and the state-transition probabilities ($p$, $r$, $k$ and $h$) when a Gilbert-Elliott loss model is used. The parameters values are inspired from the work of Rula \textit{et al.} on In-Flight Communications~\cite{inflightconnectivity}. As the authors specified one representative set of parameters for \textit{Direct Air}-\textit{to}-\textit{Ground Communication} (DA2GC) and one set for \textit{Mobile Satellite Service} (MSS), we built our ranges around these values and perform experiments with many combinations of different values within these ranges, covering most of the loss conditions they experienced. For each set of parameters, the experiment uses 4 file sizes: 1kB, 10kB, 50kB and a larger file of 1MB. These sizes are intended to represent typical file sizes for the web browsing use-case.

Unless otherwise specified, the level of redundancy is set to $(30, 20)$ for the Reed-Solomon code and (3, 2, 20) for the RLC FEC Scheme. This ensures a code rate of $\frac{2}{3}$ and a burst recovery capability of 10 symbols per block.

\begin{table}
\centering
\begin{tabular}{|c|cccccc|}
\hline
Parameter & bw (Mbps) & $p$ & $r$ & $k$ & $h$ & OWD (ms)\\
\hline
Smallest & 0.3 & 0.01 & 0.08 & 0.98 & 0   & 100 \\
Highest & 10 & 0.08 & 0.5  & 1    & 0.1 & 400 \\
\hline
\end{tabular}
\caption{Experimental design parameter ranges.}
\label{table_parameters_range}
\end{table}

\subsubsection{Reproducible experiments}

We advocate for having reproducible experiments in order to easily analyse the results. It also enables a fair comparison between the two solutions and assess them under equal conditions. Mininet already provides tools to emulate uniform losses on a link but neither provides a Gilbert-Elliott loss model nor a way to reproduce the loss pattern of an experiment. We thus built our own tool, \texttt{ebpf\_dropper}, allowing emulating losses in a network. 
This tool, written using the \textit{extended Berkeley Packet Filter} (eBPF) \cite{ebpf}, can be attached to a network node via the \texttt{tc} tool. It provides a uniform and a Gilbert-Elliott deterministic loss model, which can be given a seed to exactly reproduce the sequence of lost packets.

\subsection{Results with uniform losses}

As Rula \textit{et al.}~\cite{inflightconnectivity} proposed a uniform loss rate for the IFC use-case, we first perform experiments with uniform losses and investigate the benefits of FEC in QUIC in these configurations. We first study the two average cases for IFC. We then extend the parameters ranges using experimental design. We compare the regular QUIC with our QUIC-FEC implementation, using different error-correcting codes: the RLC and Reed-Solomon codes. We do not present the results for the XOR code, as it showed similar or poorer results to these two codes. 

\subsubsection{Specific IFC use-cases}

In this section, we study in details the average parameters values proposed by Rula \textit{et al.}~\cite{inflightconnectivity} for Mobile Satellite Service and Direct Air-To-Ground Communications with different deterministic uniform loss patterns.
For each case, we performed 50 experiments for each file size with a different seed for our deterministic loss generator, allowing experimenting with a high variety of loss patterns. Due to space limitations, we do not show here the results with the Reed-Solomon code here as RLC seems to outperform it in this uniform losses environment. 

\begin{figure}
\centering
\includegraphics[width=0.9\linewidth]{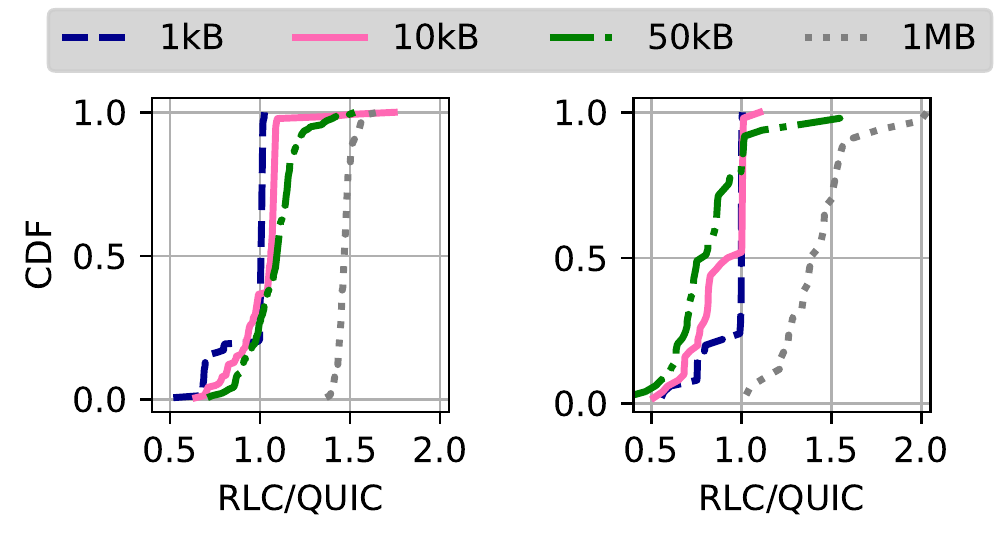}
\caption{Download Completion Time (DCT) ratio between QUIC-FEC with the RLC block code and QUIC for the average DA2GC (left) and MSS (right) parameters values. A ratio below 1 means that QUIC-FEC performed better than QUIC. For DA2GC, FEC was only beneficial for 25\% of the experiments for small file sizes. FEC noticeably deteriorates the performance in the other cases, due to the low bandwidth. For MSS, FEC can improve the performances in 75\% of the cases for 50kB file transfers and does not deteriorate it when no loss occurs during small file transfers, due to the higher available bandwidth.}
\label{figure_results_average_cases}
\end{figure}

\paragraph{Direct Air-To-Ground Communication (DA2GC)}
We experiment with the average parameters values for the DA2GC use-case~\cite{inflightconnectivity}. 
This leads us to a Round-Trip-Time of 262ms, a link bandwidth of 0.468 Mbps and a loss rate of 3.3\%. The results are presented on the left graph of Figure \ref{figure_results_average_cases}. 
As we can see, even within the same set of parameters, the experiment can lead to quite different results when run with different loss patterns. This is because the position of the lost packets in the packet flow can have a high impact on the final result. 
With small file transfers, only a few packets will be lost, if any. 
If a loss occurs on a packet transporting non-stream frames, its impact on the download completion time will likely be negligible or non-existent if the files are small (with small files transfers, the sender will not necessarily be flow control-blocked, thus losses of packets containing frames updating the flow control window won't have an impact on the file transfer). 
On the other hand, losses that occur on packets containing Stream frames can have a high impact on the download completion time if FEC is not used or did not recover the packet.

As we can see, the 1kB file download sees only a positive impact or no impact when FEC is used. 
The file is indeed not large enough to saturate the sender's congestion window even when FEC is used. However, when the packet containing the Stream frame is lost, QUIC-FEC will be able to recover it as soon as the FEC frame is received (which has been sent directly after the Stream frame, in a separated packet). 
The regular QUIC implementation will have to wait for at least one RTT to retransmit the frame. 
In the case of a 1kB file download, the loss of the only Stream frame is also a tail loss, that will be retransmitted at least after waiting $2*RTT$ in the \texttt{quic-go} implementation that uses the \textit{Tail Loss Probe (TLP)} mechanism \cite{tcp-loss-probe}. 
In this case, the sender will wait for more than 500 milliseconds before retransmitting the frame.

The advantages of FEC are less evident for larger files: for both 10kB and 50kB files, using FEC clearly deteriorates the DCT in most cases. This result can easily be explained. With such a low bandwidth, the network forwards one packet of 1200 bytes every 20.5 milliseconds. 
Even in the case of a limited number of packets such as with the transfer of a 10kB file, the impact of the additional bandwidth required to transfer the Repair Symbols will be noticed with such a limited bandwidth. 
The advantage of FEC is especially visible when a packet loss occurs during our tests: recovering the packet through FEC reduces the DCT compared to a retransmission.

\paragraph{Mobile Satellite Service (MSS)}

We performed experiments with the average parameters values for the Mobile Satellite Service use-case provided by Rula \textit{et al.}~\cite{inflightconnectivity}. 
We thus set a Round-Trip-Time of 761ms, a link bandwidth of 1.89 Mbps and a loss rate of 6\%. 
The graph at the right of Figure \ref{figure_results_average_cases} shows the DCT ratio between QUIC-FEC with RLC and QUIC without the FEC extension. As we can see, using FEC reduces the total DCT in the vast majority of the smaller files downloads. The RTT and loss rate are sufficiently high to have a highly negative impact on the DCT that will be recovered through the use of FEC. 
As the bandwidth is a lot higher than for the DA2GC case, the negative impact of FEC on smaller files is less present. Using FEC takes the benefits of the additional available bandwidth to transmit the redundancy needed to recover from losses. 
It can be easily seen when comparing the 10kB curves for the DA2GC and MSS cases: when no loss occurred, the ratios are significantly closer to 1 for the MSS case.
 Finally, we can note a higher variance of the DCT ratio for the 1MB files compared to the DA2GC scenario. This is due to the higher available bandwidth. 
With a higher available bandwidth, there is a larger set of possible values for the congestion window after encountering the first random loss making the sender exit from the slow start phase. 
The position of the first loss in the loss pattern has thus a higher impact than with a smaller available bandwidth.

\subsubsection{Experimental Design}
\begin{figure}
\centering
\includegraphics[width=0.9\linewidth]{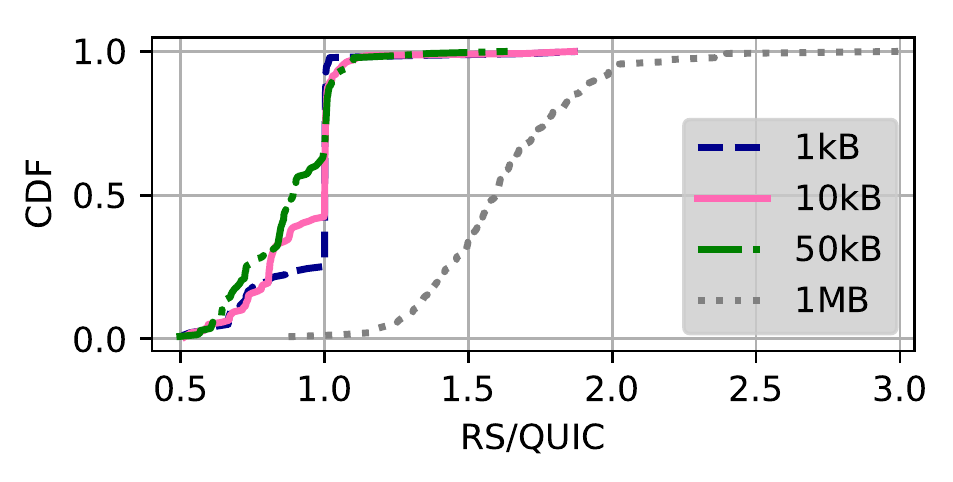}
\caption{DCT ratio between QUIC-FEC with the Reed-Solomon block code and the regular QUIC.}
\label{figure_results_rs_vs_quic}
\end{figure}

 We now use the Experimental Design approach to explore a broader set of parameters values. Figure \ref{figure_results_rs_vs_quic} shows the ECDF of the ratio of the Download Completion Time (DCT) between QUIC-FEC using Reed-Solomon and the regular QUIC, with four different file sizes. Each experiment in this CDF has been performed with parameters selected from the ranges shown in Table \ref{table_parameters_range}. 
\paragraph{Large files transfers}
We can easily see that QUIC-FEC performs badly compared to the regular QUIC with the 1MB file. This comes from one of the drawbacks of FEC. Adding redundancy to a data stream implies to transfer more bytes on the wire. As our code rate is $\frac{2}{3}$ in this experiment, QUIC-FEC has to transfer 1.5 times the amount of bytes transferred by regular QUIC (ignoring the potential retransmissions), which increases the overall download completion time. With longer files, the benefits brought by FEC are thus masked by the overhead needed to transmit the redundancy. In some cases, we can see that the DCT ratio can reach high values when the first losses arrive later. This will let the congestion window increase higher and the regular QUIC download will be able to terminate before the congestion window finished to decrease to a stable level. On the other side, QUIC-FEC will have to transmit the remaining data at a smaller rate and thus be significantly slower.

\paragraph{Small files transfers}
We can see that the advantage of FEC is visible on small file transfers. 
Recovering a packet with FEC avoids the wait for a retransmission. 
A retransmission costs at least the time of one Round-Trip-Time. 
With small files, the wait for a retransmission will have a high relative impact on the DCT. 
This is why using FEC has an advantage in these cases, as the file transfer will be able to terminate without an additional round trip. 
The experimental design also shows that there is a benefit in protecting the client request. 
For example, our 108th test discards the client packet containing the GET request. 
The server then recovers it without the need for the client to retransmit it and can directly begin to serve the request. 
However, it should be noted that protecting the client's request is equivalent to duplicating it if the request is contained in only one packet.

When looking more closely at our results, we can also see that FEC can still be harmful, depending on the loss pattern and the available bandwidth. 
Indeed, for experiments during which no loss occurred and with a low available bandwidth, using FEC sensibly increases the DCT, even for the 10kB and 50kB files transfers. These configurations are indeed similar to the DA2GC scenario. The additional time needed to transmit the redundancy is non negligible compared to the total DCT. When losses occur or for experiments having a higher available bandwidth, this overhead is greatly reduced.

\paragraph{Comparing FEC codes}

\begin{figure}
\centering
\includegraphics[width=0.9\linewidth]{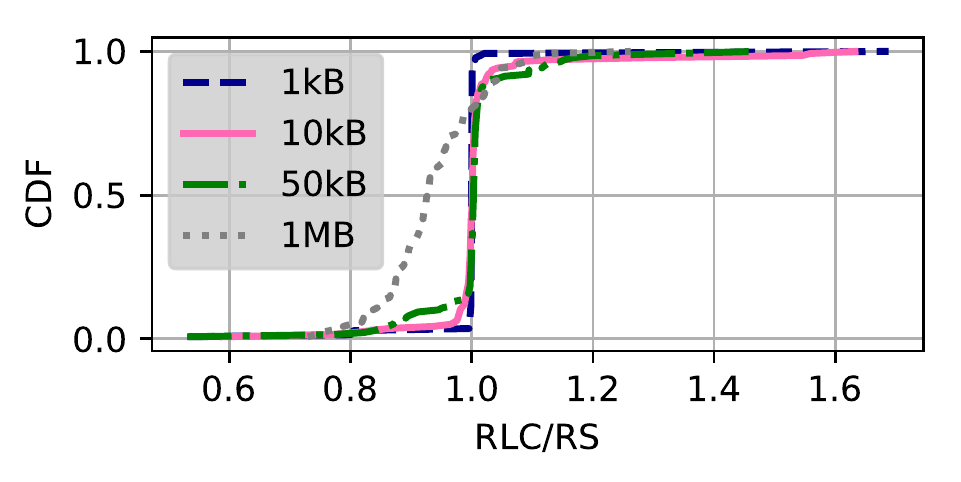}
\caption{DCT ratio between QUIC-FEC with the convolutional RLC code and QUIC-FEC with the Reed-Solomon block code.}
\label{figure_results_rs_vs_rlc}
\end{figure}

We now compare the impact of the FEC Scheme used for these different file sizes. Figure \ref{figure_results_rs_vs_rlc} shows the DCT ratio between QUIC-FEC using the Reed-Solomon block code and QUIC-FEC using the RLC convolutional code. As we can see, RLC performs significantly better than Reed-Solomon with the 1MB file transfer. This can be explained easily by the way these two codes send their Source and Repair Symbols. These two codes provide the same code rate and a similar packet recovery capability. However, the RLC code interleaves the packets containing the FEC frames with the FEC-protected QUIC packets. On the other hand, the Reed-Solomon code sends all its FEC frames after the block containing the FEC-protected packets has been sent. In our experiments, the RLC code sends one FEC frame every two FEC-protected packets. The Reed-Solomon code sends 10 FEC frames every 20 FEC-protected packets. If the first FEC-protected packet of a Reed-Solomon block is lost, the receiver will have to wait for receiving the 19 other FEC-protected QUIC packets and one Repair Symbol before being able to recover it. On the other hand, the RLC code must only wait for the reception of two additional symbols: the following QUIC packet and the following FEC frame. With a packet-based loss detection threshold of 3 packets such as the one defined in the QUIC recovery draft \cite{quic-rec-draft-16}, using the Reed-Solomon code will trigger a retransmission on the sender in most cases as the packet has been recovered too late. The uselessly retransmitted packet will occupy the congestion window of the sender, while a new packet will be sent when RLC is used.

\subsubsection{The need for adaptive FEC}
\begin{figure}
\centering
\includegraphics[width=0.9\linewidth]{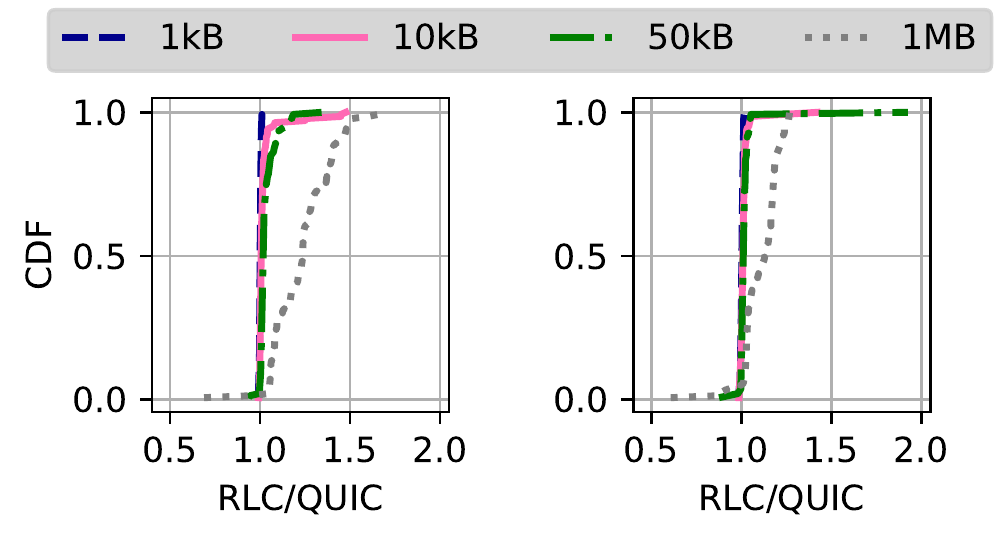}
\caption{DCT ratio between QUIC-FEC with RLC and the regular QUIC with a loss rate of 0\%. On the left, the used code rate is $\frac{2}{3}$, while on the right, the code rate is $\frac{4}{5}$. Increasing the code rate reduces the impact of FEC on the bandwidth.}
\label{figure_results_rlc_vs_quic_no_loss}
\end{figure}

Using the experimental design, we saw that FEC performed badly when no loss occur and when the available bandwidth is low. 
We want to study to which extent using FEC can deteriorate the DCT.
We perform experiments with the same parameters, except for the loss rate that we set to 0\%. 
Figure \ref{figure_results_rlc_vs_quic_no_loss} shows the DCT ratio comparing QUIC-FEC with RLC and regular QUIC. On the left (resp. right) of the figure, QUIC-FEC uses RLC with a code rate of $\frac{2}{3}$ (resp. $\frac{4}{5}$).
As we can see, except for some results due to a slight variance in our experiments, using FEC always deteriorates the DCT. 
When looking at our results, we see that the DCT is further deteriorated during experiments for which the available bandwidth is low. When comparing the results with the different code rates, we can see that increasing the code rate reduces the negative impact of FEC on the DCT. We can thus conclude that there is a need in controlling the redundancy level during the life of a connection and disabling it when it deteriorates the DCT.

Adaptive coding schemes are already studied in the literature at different levels in order to reduce the negative impact of over-coded transmissions~\cite{nafaa2008forward,hou2008adapcode,
zhang2004channel,chaporkar2007adaptive,zhang2005end,cloud2015coded}. It is however known that ARQ offers better performances than FEC for larger bulk transfers~\cite{zhang2005end}. This is also what we can observe from our experiment results. Disabling the use of FEC for these use-cases should thus be considered.

\subsubsection{The importance of the recovery notification}
\label{subsection:recovered}

In Section \ref{section_fec_congestion} we proposed the Recovered frame to notify the peer that packets have been recovered to avoid hiding the loss-based congestion signal. In this section, we analyse the impact of this notification on the fairness of QUIC-FEC.

The connection parameters used for this experiment are the parameters of the MSS scenario, except that we set a loss rate of 0\% to avoid disturbing the experiment with random losses. Losses will only be caused by congestion. We use a $(7, 6, 20)$ RLC code, leading to a code rate of $\frac{6}{7}$, to better see the impact of masking the congestion signal when no Recovered frame is sent. Indeed, with a higher code rate such as $\frac{2}{3}$, a large part of the packet flow is composed by packets containing only a FEC frame. The loss of a single FEC frame does not lead to the transmission of a Recovered frame since this loss does not impact the receiver, but it is still announced in ACK frames.
The sender thus still receives a frequent loss signal when many FEC frames are lost. With a higher code rate such a $\frac{6}{7}$, losses of packets containing only FEC frames and the related congestion signal will be less frequent. Note that \texttt{quic-go} also marks a packet as lost if there is a hole in its ACK ranges for this packet for more than $\frac{1}{8}*RTT$.

\begin{figure}
\centering
\includegraphics[width=0.9\linewidth]{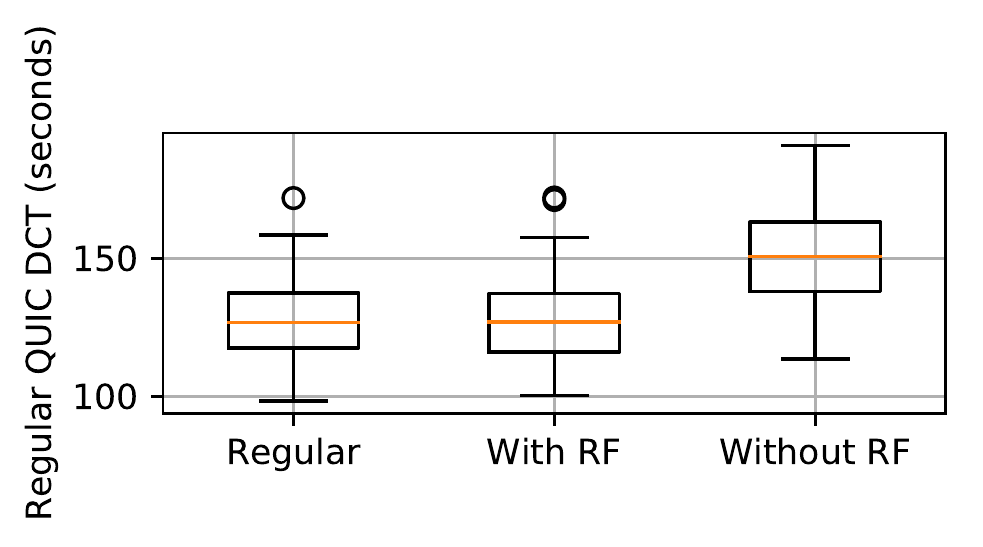}
\caption{DCT of a regular QUIC transfer when competing with QUIC, QUIC-FEC with Recovered frames and QUIC-FEC without Recovered frames.}
\label{figure_fairness_comparison}
\end{figure}

Our experiment consists in performing a 10MB file transfer with the regular QUIC implementation (we call it the \textit{foreground transfer}) while the link is already fully utilised by another transfer (we call it the \textit{background transfer}). 
We consider three different candidates for the background transfer: $1)$ another regular QUIC transfer, $2)$ a QUIC-FEC connection that uses Recovered frames (RF) and $3)$ a QUIC-FEC transfer that simply acknowledges the recovered packets, without sending Recovered frames. We compare the DCT of the foreground transfer in these three cases. The left, middle and right box plots in Figure \ref{figure_fairness_comparison} respectively represent the DCT of the foreground transfer for the first, second and third cases.

As we can see, the regular QUIC download takes generally longer when it competes with a QUIC-FEC transfer that does not send Recovered frames. This is due to the fact that in this case, the congestion signal is lost when packet losses caused by congestion are recovered and simply acknowledged by the receiver. This makes a FEC-enabled protocol unfair compared with traditional protocols such as regular QUIC that only use retransmissions. The middle bar plot shows that when QUIC-FEC uses Recovered frames, there is no difference between a QUIC sender competing with another QUIC sender or a QUIC-FEC sender.

\subsection{Results with bursty losses}

\begin{figure}
\centering
\includegraphics[width=.9\linewidth]{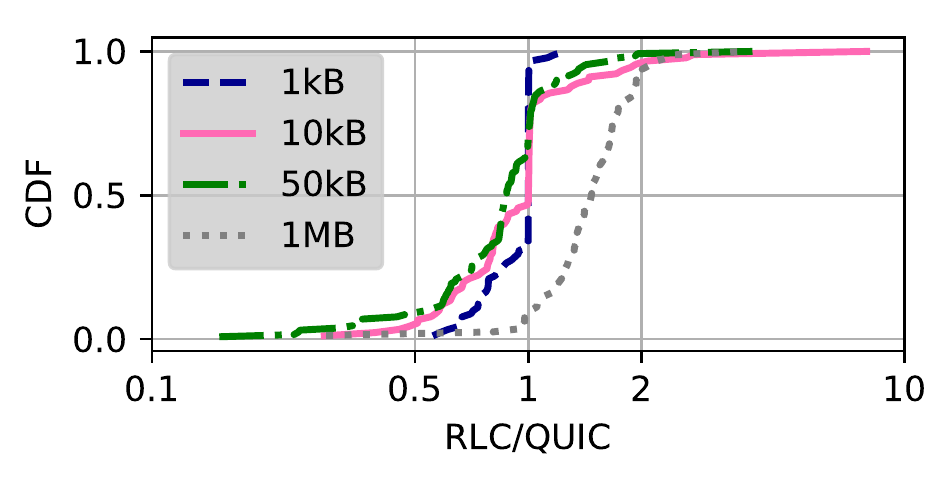}
\caption{DCT ratio between QUIC-FEC with the RLC block code and QUIC, using the Gilbert-Elliott loss model.}
\label{figure_results_rlc_vs_quic_gemodel}
\end{figure}

In this section, we analyse the impact of using FEC in the case of correlated losses. 
We perform the same experiments with a Gilbert-Elliott loss model. 
The parameters of these experiments are shown in Table \ref{table_parameters_range}. We remove from our results the experiments whose intense loss patterns prevent a successful file transfer. We show the comparison of QUIC-FEC with RLC and the regular QUIC in Figure \ref{figure_results_rlc_vs_quic_gemodel}. 
We can see that the results are close to the results with uniform losses shown in Figure \ref{figure_results_rs_vs_quic}: using FEC has benefits for smaller transfers and the 1MB file transfer suffers from the added redundancy.

When looking more closely at our results, it appears that using FEC performs badly when the $r$ parameter of the Gilbert-Elliott model is low. For the download of 10kB and 50kB files, the average DCT ratio for our experiments when $r \leq 12\%$ is above 1, while it is below 1 otherwise.

\subsection{Experiments conclusion and future work}

As a first conclusion, we can note that adding FEC to a file transfer considerably reduces the Download Completion Time for small files. For such transfers, a loss will more likely impact the Download Completion Time as its retransmission will potentially occur after all data have been sent. Using FEC avoids waiting this additional time. FEC is beneficial in higher-bandwidth networks. The MSS use-case already proposes a sufficient bandwidth to hide the overhead induced by FEC for smaller file sizes. Newer MSS and DA2GC technologies aim at improving the available bandwidth~\cite{inflightconnectivity}, and thus the efficiency of FEC.  We compared a block to a convolutional code and showed that the convolutional code can recover from single loss events more rapidly than block codes, leading to a better bandwidth efficiency by avoiding useless packet retransmissions for long transfers.

We can also conclude that using FEC requires to be able to adjust the code rate, or even disable the transmission of redundancy, especially for large file transfers.

\section{Conclusion}
In this paper, we have leveraged the extensibility of QUIC to complement its existing retransmission techniques with Forward Erasure Correction (FEC). FEC is particularly suited for lossy communications over high-delay paths such as In-Flight Communications.
Our design and implementation are modular and can support a variety of FEC techniques with different levels of redundancy. Furthermore, thanks to our Recovered frame, the sender can correctly adjust its congestion window when the receiver uses FEC to recover from packet losses.

Our evaluation over a wide range of network scenarios shows that FEC can drastically lower the Download Completion Time for short web transfers. However, it can be harmful for longer file downloads or in low-bandwidth configurations. Our future work will be to develop heuristics that enable QUIC senders to automatically decide whether to use FEC or traditional retransmission techniques in function of the requirements of the application and the network conditions. We also plan to use our implementation to carry out large scale experiments in real networks.

\section*{Artefacts}
Our code for QUIC-FEC and the scripts for the experiments\footnote{Available on this link: \url{https://bitbucket.org/michelfra/quic-fec} 
on the branch \texttt{networking\_2019} from May 20th 2019. } as well as \texttt{ebpf\_dropper}\footnote{Available on this link: \url{https://github.com/francoismichel/ebpf_dropper} 
on the branch \texttt{networking\_2019}.} are publicly available.

\bibliographystyle{IEEEtran}
\bibliography{IEEEabrv,reference}
\end{document}